Knowledge Diffusion Process & Common Islamic Banking Governance Principles: Integrative Perspective (s) of Managers and Shariah Scholars

**Adnan Malik**

Ph.D. Scholar, Institute of Management Sciences, Peshawar

**Dr. Karim Ullah**

Assistant Professor, Institute of Management Sciences, Peshawar

**Dr. Shakir Ullah**

Assistant Professor, Institute of Management Sciences, Peshawar

**Abstract**

Islamic banks being commercial entities strive to earn profit within shariah ambit. Therefore, they seem to be basing themselves upon two knowledge streams namely i) Islamic jurisprudence principles, and ii) banking principles. Islamic jurisprudence principles primarily aim at bringing shariah compliance while banking principles focus profitability. These principles, making two schools of thought in the discipline, however, have their unique philosophies, principles, and practices, which are now gradually diffusing into an emergent set of governance principles basing the contemporary Islamic banking theory and practice. Governance systems of Islamic banks have elements of both conventional as well Shariah, and need to have principles having components of banking and shariah sufficiently diffused for their successful operations in a longer term. Aim of this research is to review the literature about the knowledge diffusion process of islamic banking principles which guides the governance of Islamic banks. This study review the literature using a method in which focus remain on bridging different areas which in this case are knowledge diffusion and islamic banking governance principles.

The background theory of Islamic banking reveals that it is basing upon the Shariah financial principles of avoiding interest, uncertainty and promoting profit and loss sharing in transactions. Whereas, the conventional banking is, primarily, informed by the principles of profitability, solvency, and liquidity. The research adopts knowledge, aptitude, and practice (KAP) as focal theory to establish an initial pre-empirical framework, that put-up three propositions as: i) The interpretations of Shariah are adopting banking principles ii) The interpretations of Banking principles adopting Shariah principles iii) There are emergent diffused governance principles basing the current Islamic banking practices.

*Corresponding Author: Adnan Malik, Lecturer & PhD Scholar Institute of Management Sciences, Phase 07 Hayatabad, Peshawar, Pakistan.*

*Adnan Malik is a certified professional in insurance and takaful with nearly 15 years of practical experience in reputable life and general insurance and takaful companies. Since 2011, he is teaching insurance and takaful related subjects at Institute of Management Sciences, Peshawar. Adnan is the author of book "Introduction to Takaful; Theory and Practice" published in 2019 by UK based, world leading publisher "Palgrave Macmillan".*

*Cell Number: 00923005881809*

*adnan.malik@imsciences.edu.pk*

## 1. Introduction and background of the Study

Islamic financial industry has witnessed steady growth, since its re-emergence in 1970's in Middle Eastern part of the world. This growth, is three dimensional, as Islamic finance is growing in new product-lines, sectoral, and geographical (Ullah & Patel, 2011) with estimated asset base of $6.7 trillion by 2020 (S&P, 2018) and more than 500 Institutes worldwide (Qorchi, 2005). Firstly, Islamic banks were offering few products based on musharika, modarbah, murabaha, and ijarah (Othman and Mara, 2013) but now, they are offering more products, based on other contracts like salam, istisna, running musharika and wakala. Secondly, in 1970's, only islamic banks were the institutions established on shariah principles but later on, other institutions related to takaful, mutual funds, sukuks, islamic microfinance, modarbah companies, shariah audit and screening companies also established (Kahf, 2006). Thirdly, in 1970's islamic financial institutions were concentrated mainly in middle east but now, operating in over 60 countries including USA, Australia, Russia, Germany, New Zealand (Iqbal, 1997; Kuran, 2004, Elasrag, 2014). This growth seems to be happening due to diffusion of knowledge principles of shariah with banking, insurance, microfinance and investment companies.

This multi-dimensional growth is, primarily, motivated by the rigorous research and knowledge of Islamic jurists and economists(Bhala, 2013). Various learning and research centers were established for the promotion of Islamic economy in Saudi Arabia, Pakistan, Egypt and Malaysia (Kuran, 2004).Since, islamic banks claim to follow shariah rules in operations, therefore, such institutional structures becomes inevitable which ensures governance and operationalization of shariah in practices (Ullah, Harwood and Jamali, 2016). Therefore, institutions like Accounting and Auditing Organization for Islamic Financial Institutions (AAOIFI) and Islamic financial Services Board (IFSB) are established for developing standards for promoting harmony and standardization in islamic financial industry (Kasim & NuHtay, 2013). AAOIFI was established in 1991 for developing standards for global Islamic financial institutions. They have so far developed more than 100 standards relating to shariah, accounting and auditing standards (AAOIFI, 2018). IFSB was established in 2002 at Malaysia. It develops regulatory standards for the stability of Islamic financial industry. It has so far issued 27 standards on various areas including risk adequacy, risk management and solvency requirements (S&P, 2018).Around 45 countries have islamic banking regulations and they mostly takes guidance in developing their shariah governance frameworks from AAOIFI and IFSB (Thompson Reuters, 2018).

Recent practices and extensive literature is talking about fundamental and managerial issue of clashes and misunderstandings among shariah scholars and managers working in islamic banks (Ullah, 2012; Haridan, Hassan, & Karbhari, 2018). Shariah scholars do not understand the essential business principles and value system of the banking while the managers do not understand the essential principles and of shariah and value system of shariah scholars. These misunderstandings are due to the lack of knowledge about each other, is leading to the confrontations (Rammal, 2015;Wilson, 2000; Farook, 2011; Hamza, 2013). These confrontations are discussed in the literature. The core principles of these two schools of thought are relatively less discussed in the literature.

Islamic banks are operating since 1970's generating profits for their shareholders and customers in a shariah compliant way. There practices are different than a conventional bank. Still they are working successfully and seems that they will operate in a longer term. A financial institution if running successfully should have its own governance principles translating into practices. Can an Islamic bank generate profit in a shariah compliant way following only banking principles or only shariah principles of governance? If no, then what are the common principles which inform the governance of Islamic banking? Perhaps, a new set of principles informing shariah and banking practices may have developed containing elements of shariah and banking principles through cross knowledge diffusion among managers (conventional bankers) and shariah experts.

Presently, more than 500 Islamic banks are operating globally in 56 countries and has 1.72 trillion US$ assets which makes 6% of total the banking. 1162 shariah scholars are working in the islamic banking (Thompson Reuters, 2018). It seems that while working together in Islamic banks, shariah scholars and managers may be exposing to each other's knowledge and diffusion of shariah and banking knowledge may have occurred. At times, conflicts also arise among them which are reported in the literature but since Islamic banks are operating successfully. Therefore, the conflicts may be settling perhaps due to increase in understanding each other's value propositions. The understanding may be increasing due to diffusion of shariah and banking knowledge making them hybrid professionals.The diffusion may be resulting in developing common practices and principles which both experts follow. These common governance principles of Islamic banking if exist, are not known and needs to be explored.

This study will try to explore the literature on knowledge diffusion, contributing into emergence of commonIslamic banking governance principles.

## 2.    Literature Review Methodology

Literature is reviewed with different purposes. It may be assessing current knowledge level, future research directions and guiding decision making (Guzzo, Jackson and Katzell, 1987). Through Creswell (2014), Cooper (2010) states that literature review is carried out with any of the four purposes either (a) integrating others work b) critically evaluating others work (c) bridging related areas and (d) identifying central issues in a field.

This study reviews literature for bridging the knowledge diffusion process and Islamic banking governance principles. Literature on Knowledge diffusion and governance in Islamic banking is available in isolation but lesser work seems to be carried out on how knowledge diffusion is contributing into the emergence of Islamic banking governance principles. Literature, however, reports instances where knowledge is diffusing among people working in Islamic banks and emergence of practices having the impact of governance in Islamic banking different than conventional banking. Therefore in this study, literature on knowledge diffusion and governance is reported in detail focusing the link between these areas.

Literature is reviewed using the following methodology in order to make it comprehensive. Relevant papers with words knowledge diffusion, governance, Islamic banking were searched using the science web. Other sources like ProQuest, Google scholar; Elsevier, JSTOR, Emerald etc. were also searched while using the above mentioned words. Most of the reviewed papers are those published after 2010 but the oldest one published in 1994 was also considered. Backward and forward referencing checking was also carried out so that more and more relevant papers are included in the study. For further enriching the literature, relevant books of seminal authors and reports of reputable professional bodies were also considered. The filtered research papers, books and reports were used as documents and analyzed using one of qualitative research techniques which is thematic analysis. After filtering, around 75 documents including research papers published in reputable peer reviewed journals, books, and industry reports were selected. 47 are research papers while the rest are industry reports and books included in the final sample. Themes in the form of Shariah financial principles and banking principles were identified and a literature based framework was developed using them.

**3.    Literature Review**

### 3.1. Knowledge

Knowledge is a complex phenomenon and multiple perspectives exist about it. Famous Greek philosophers also try to define it. Socrates claims that humans already know things but exploring that knowledge is only recollection of it (Baronett, 2016). Aristotle keeps a different view and declares experience as the source of knowledge (Bronstein, 2016) which is more close to interpretation with the today's scientific knowledge acquired through reason and senses.Plato considers knowledge as a justified true belief (Nagel & Mar, 2013) while Nonaka (1994) adds further to this perspective that being a belief, knowledge enhances one's capacity for taking an effective action. Nonaka (1994) suggests a model showing conversion of tacit knowledge into expressed knowledge through four modes.

This study takes the concept of knowledge given by Fahey and Prusak (1998) which say that knowledge lies within the knower and new knowledge creates when new knowledge mixes with the existing knowledge. Knowledge may not be transferring but is created in the receiver's mind as transferring anything reduces with the sender but knowledge remains at the same level with the sender after passing on to the receiver.

### 3.2. Diffusion

Merriam Webster dictionary ( 2019) defines diffusion as "the spread of cultural elements from one area or group of people to others by contact". "Diffusion of Innovation Theory" developed defines it as "Diffusion is a process in which innovation is communicated through certain channels over time among the members of a social system"(Rogers, 2010).  This innovation can be a new idea, product or any human behavior.

Islamic banking started in 1970's was a new and emergent phenomenon. Research in this area was taking place from a considerable time while new institutions started establishing in 1970's on the idea of doing business without having the presence of interest, uncertainty following Islamic financial principles. So, people well versed in shariah (Islamic law) teamed up with managers (people having banking knowledge) developed a banking system which was following shariah and was profitable as well. They managed to build a knowledge base having elements of shariah financial principles like avoiding interest, uncertainty and promoting profit and loss sharing (Ilieva, Ristovska and Kozuharov, 2017)along with keeping the bank solvent, liquid and profitable. It seems that Knowledge streams of shariah and banking combined when shariah scholars and managers worked together. Initially, both were following their own perspectives. Shariah scholars were only looking for shariah compliance i.e. they were not aiming at market viability of the product but were focusing the absence of interest and uncertainty.Similarly, managers were only focusing their banking principles of being profitable, solvent and liquid. Not knowing each other value propositions resulted in confrontations (Ullah, 2012)which are reported in the literature. Despite of confrontations, still many Islamic banks are operating successfully in a shariah compliant way, yet profitable. It seems that cross diffusion of banking and shariah knowledge is happening among shariah scholars and managers respectively.

Focal theory for this research is KAP theory which stands for Knowledge, attitude and practice. "KAP" studies are carried out for finding the knowledge level of a phenomenon and its practice in a community (Kaliyaperumal, 2004). It can also identify an intervention for bringing more awareness and increasing the practice of the phenomenon. A knowledge, Attitude and Practice (KAP) survey conducted by Mehtab, Zaheer, & Ali (2015) showed that knowledge has a strong relationship with practices of Islamic banking. One of such KAP study was conducted in 2014 by Central bank of Pakistan which finds that more than 95% people in Pakistan desire to avoid interest and 74% of conventional bank customers want to switch to Islamic banking. 65% religious sensitive people doubt the shariah compliance of existing Islamic banking, although, they don't have sufficient knowledge about it (KAP study, 2014). KAP theory says when someone acquires knowledge about a phenomenon; he thinks about it with his existing knowledge and develops an attitude accordingly. The developed attitude may be positive or negative about the phenomenon. If it is positive and the person starts thinking differently about the phenomenon, perhaps, the belief has changed. This change in belief results in different practice.

People receiving knowledge of economics, when join banking, they also bring their beliefs regarding business. They are taught that business exist for a profit  and when it comes to banking business, liquidity and solvency are the banking fundamentals that needs to be observed. At this stage, when they asked to avoid interest and uncertainty while promoting profit and loss sharing transactions, they may not comprehend and may claim that they are doing banking for a profit only.

People gaining knowledge of shariah in religious schools develops their beliefs (principles) on compliance and accountability before God. They develop their attitude of avoiding prohibited elements in banking like interest, uncertainty and gambling. So, when they join an islamic bank, they try to follow their principles and hence, a difference emerge among bankers and shariah experts which sometimes lead to a struggle for dominance.

Knowledge of Islamic banking is diffusing among the people and several institutions are playing their roles in it. Globally, 688 institutes are providing Islamic finance education. Malaysia and Pakistan are the top education providers. In 2015 – 2017, 2564 research papers on islamic banking were published in scientific journals. In 2017, more than 417 awareness sessions were conducted on islamic banking and 13257 news items were published. This shows that a complete ecosystem is developing for the grounding of this phenomenon among people (Thompson Reuters, 2018).

### 3.3. Governance (CG)

CG existed as a practice since companies were formed while as a study; it gained momentum in 1970's onwards when problems in governance occurred in companies operating in US, UK and other parts of the world. Large frauds occurred in big companies like Enron, WorldCom, Polly peck international, Barings bank, Bank of Credit and Commerce International (BCCI), which led to the thought of improving governance structures(Yu, 2015; Majdoub, 2015).Organization for Economic Cooperation and Development (OECD) and professional bodies like Association of Chartered Certified Accountants ACCA developed codes of governances. Various countries like UK (UK Code of Governance in 1998), USA (Sarbanes Oxley Act 2002), and Australia etc. made laws for improving governance in public limited companies for avoiding frauds and improving confidence of investors (Butt, 2010). Focus increased further when the recent recession of 2007 started and large banks like Lehman Brothers collapsed.

Islamic banks also need governance as several islamic financial institutions also collapsed, like in 1997, the closure of Islamic bank of South Africa, the collapse of Turkey based company, Ihlas Finance in 2001 are the few examples. Therefore, IBs also need to have efficient governance systems also catering the need of shariah governance(Grassa, 2018). Since, islamic banks claim to follow shariah rules in operations, therefore, such institutional structures becomes inevitable which ensures governance and operationalization of shariah in practices (Ullah, Harwood and Jamali, 2016).

For supplementing governance standards for IFI's, organizations like AAOIFI, IFSB make standards which keep the unique feature of Shariah compliance. International bodies like AAOIFI and IFASB are promoting standardization within global islamic financial industry which contributes in better shariah compliance and profitability for the companies (Elasrag, 2014;Hassan, et al., 2017). IFSB defines the term "Shariah Governance System as it refers to structures and processes adopted by stakeholders in the IFSI to ensure compliance with Shariah rules and principles"(IFSB, 2009). Experts of banking and shariah, both, work together at these forums and collectively contribute in developing standards for Islamic banking industry aiming at improving shariah compliance and growth in the sector. The cross diffusion of shariah and banking knowledge among the experts is contributing in the islamic banking growth with more than 500 islamic banks having USD 1.72 trillion US$ are operating globally (Global Islamic Finance Report, 2019).

In Islamic banks, in addition to the Board of Directors (BOD), there exist Shariah supervisory board (SSB) which is entrusted with the task of enforcing shariah compliance within banking.Being companies, Islamic banks mention their operations to be shariah compliant within articles and memorandum of association which makes it a fiduciary duty for the BOD to ensure shariah compliance within the bank(Othman et al., 2013). IFSB issues standard for shariah governance which requires the BOD to give independence to SSB in its decisions and meet with it on regular basis.Meeting of the members of BOD and SSB also encourage making of policies addressing shariah compliance and profitability needs. Such meetings furthers diffuses the banking and shariah knowledge.

Shariah Supervisory Board is considered to be an important set up within Islamic banks for ensuring shariah compliance (Hamza, 2013). Presence of SSB in Islamic banks gives a peace of mind to the customers that the services and benefits they are receiving are shariah compliant. BOD normally follows banking principles while SSB implements shariah principles within the banking (Chowdhury & Shaker, 2015). AAOIFI Governance Standards No. 1 & 2 for Islamic banks defines "Sharia Committee" as "it is an independent body entrusted with the duty of directing, reviewing and supervising the activities of the Islamic financial institution in order to ensure that they are in compliance with Islamic Shariah rules and principles". It also mentions the need of at least three members in the SSB. It issues shariah rulings (Fatwas) of certifying or non-certifying the permissibility of products, practices and services, confirming the following of shariah in occurred transactions, deciding upon the questionable income (Hassan *et al.*, 2017).In contemporary practice, most of the IBs appoint between three and six members. In some countries, there is a decentralized system of SSB like in GCC countries and Indonesia while in Malaysia (Othman et al., 2013), Pakistan, Sudan, Central banks have their SSB which keep a check on SSB established in Islamic banks. There are different opinions about keeping SSB centralized or decentralized.

Prior to 1970's, there was no concept of banking without interest. Moreover, shariah scholars were also not in favour of banking. But later on with the commencement of Islamic banking in 1970's proved that banking without interest is possible and shariah scholars also started believing in banking. All this happened due to sufficient cross knowledge diffusion of banking and shariah among shariah scholars and bankers respectively.

In rules, SSBs are given sufficient powers for bringing shariah compliance within Islamic banks. However, in practice, they face several challenges. They are expert in shariah but have lesser understanding about banking and its complexities (Haridan, Hassan &Karbhari, 2018).

Management expects that SSBs will have an accommodating view while deciding to add a doubtful profit earned into the charity fund as it reduces the profitability of the bank as well as the customers. At times, the management forces the SSB to approve a process or product although doubtful in shariah compliance which may be resulting into more profit. Such practices converges islamic banks towards the conventional banks and the customers confidence level falls (Hamza, 2013;Chapra and Ahmed's, 2002)Rammal (2015)finds in a study that in Pakistan, mangers of conventional banks perceives Shariah scholars of their banks window Islamic banking as a "negative" entity due to differing philosophies.

Several studies are carried out about the conflicts and misunderstandings between the management and shariah scholars in Islamic banks. Grassa (2013) studied the agency problem and conflict of interest arising when a shariah scholar works on SSBs of multiple banks. Through Zahid & Khan(2019),Al Qari (2002) finds that ability of independent decision making of a shariah scholar effects with the remuneration paid for the services. Findings extend further when Grais & Pellegrini (2006) reports that conflict of interest occurs when SSB members are paid from the same business which they asses.Othman et al., (2013) points out the scarcity of shariah experts having financial skills which results in inadequate shariah compliance.

Practice of Islamic banks is primarily focussed on earning profit while avoiding interest and speculations (Injas et al, 2016). Instead of providing loans, their business is more participatory based on profit and loss sharing and earning fee against the services they provide (Dusuki, 2008;Chowdhury & Shaker, 2015). They don't aim at reducing poverty, fair distribution of wealth but earn profit while avoiding the prohibited elements. Jan (2013) and Elasrag, (2014) argue that Islamic banks were established in 1970's in Middle East for earning permissible profit on the money earned from the sale of newly explored oil. Their motive was not bringing economic justice but to earn profit free from interest and speculations. Islamic banks have the option of giving interest free loans but its practice is at minimal side. Siddique (2006) supports the same notion when he claims that large banks like HSBC, Citi bank and ABN AMRO started Islamic banking in order to prevent their religiously sensitive Arab clients from shifting their deposits in other islamic banks.

Most of the Islamic financial institutions (IFI's) are commercial entities earning profit within shariah ambit. They operate to generate profit (Bashir, 1999)in a market where customers are willing to pay an additional cost for a shariah compliant service. This aim of making more profit may sometimes come in conflict with shariah, so they may not be encouraging high influence of shariah within the institution. There is always a trickle-down effect so conflicts arising between BOD and SSB, trickles down to department and branch level. So, as an output of conflicts, if shariah is compromised, they may lose market share as customers will not trust their profit to be halal. Hamza (2013) claims that in IBs, status of SSB is advisory and has no power to enforce their opinions.

In the given scenario, the conflict of interest is evident among the managers and the shariah people, resulting in a power struggle. Managers strive for profit while shariah people try to ensure compliance with shariah. Shariah frameworks suggested by AAOIFI and IFSB give absolute authority to Shariah people for ensuring compliance shariah in business transactions while in reality, management seems to follow its goal of maximizing profit (Ullah, Harwood and Jamali, 2016). Below mentioned vision of AAOIFI also supports the perspective of the Islamic Financial Institutions (IFI's) to earn profit within shariah ambit:

"To guide IFI's market operations and financial reporting on shariah principles and rules" and "To provide IFI markets with standards and guidelines that can support the growth of the industry"(AAOIFI, 2018).

Growth of the industry comes with profit which is the ultimate aim of any commercial entity. Generating profit within shariah limits is not an easy task. At times, a conflict may occur among managers focusing profit and shariah people aiming for shariah compliance. Literature reports instances of conflicts where management of some banks link compensation of shariah scholars with the size of funds, they manage, thus putting a pressure for lenient view on shariah compliance resulting in larger size of the fund (Farook & Farooq, 2011). There is a practice in some banks offering Islamic products to take a desired fatwa from a scholar not working in that bank, although the bank has its own shariah people (Wilson, 1999).Hamza (2013) claims that there can be adverse effects for the Islamic banks which compromise shariah compliance over profits. Depositors may withdraw their deposits and people in need of financing will prefer other banks having better shariah compliance. Shariah scholars having religious schooling background strive for shariah compliance while managers more focus profitability aspect of Islamic bank.

Literature shows that upbringing and grooming of the managers and shariah scholars occurs in different environments. Education and experiences of shariah scholars and managers are different. Shariah scholars get education in religious schools with emphasis on observing shariah in all matters of life (Siddiqi, 2006). They don't work in conventional banks considering it prohibited. They join Islamic banks without having any banking experience. Their education, knowledge, experience and the regulatory framework informs to ensure shariah compliance within Islamic banks. Managers acquire education which concentrate on generating profit for the business and have no formal education in shariah. Having entirely different backgrounds in terms of knowledge and experience, when shariah scholars and managers work together in Islamic banks, they rarely agree with each other on Islamic banking practices. The outcome of not knowing each other perspective becomes a conflict and power struggle (Ullah, 2012). While working together and interacting on business propositions and shariah compliance of products, there is a cross flow of shariah and banking knowledge among shariah scholars and managers which challenges their core principles of shariah and banking respectively. Consequently, they first get awared about each other principles and starts influencing one another. With a deeper interaction, they may start accepting each other perspective which according to KAP theory starts changing their feelings and attitude about each other principles. Prolong interactions may result into a changes practice based on evolved principles of shariah and banking.

From the above discussion, it is evident that role of shariah committee in governance of Islamic banks is vital. AAOIFI Governance Standards No. 1 & 2 mentioned above says that independent body i.e. shariah committee has the duty of directing, reviewing and supervising the activities of the Islamic financial institution in order to ensure that they are in compliance with Islamic Shariah rules and principles for Islamic banks. Mostly shariah scholars are the members of Shariah committees of Islamic banks and their knowledge acquired at religious schools informs their attitude and practices. They are always strict in disallowing the prohibitions like riba, Gharrar, maisir and qimar in the banking practices. On the other hand, managers running the banking operations are guided with the education received focusing on earning profit without referring to shariah compliance. Based on different educational backgrounds, shariah scholars and managers behave differently in banking. In the next section, there is a discussion on shariah financial principles which guides the practices of shariah scholars in shariah governance of banking. Banking principles are also discussed which inform managers in governing banking.

### 3.4. Shariah financial Principles

The value proposition of Islamic banking is carrying out all activities in accordance with shariah which prohibits certain elements in financial dealings. On the top of it is avoiding prohibitions namely interest, uncertainty and practicing banking with profit and loss sharing(Ilieva, Ristovska and Kozuharov, 2017). SSBs in Islamic banks exist for ensuring compliance of all dealings with the religion of Islam. Rules, regulation and procedures of all Islamic banks revolve around avoidance of non - permissible items. We may call them shariah financial principles of banking as all islamic banking activities revolves around the notion of avoiding non permissible itemsin their dealings. Let's discuss these principles one by one:

### 3.4.1.Interest Prohibition

Islam prohibits certain tools and mechanism in financial dealings. One of the prime prohibitions is dealing in interest(Naughton, 2000; Wajdi Dusuki, 2008; Chong & Liu, 2009;Yusoff, 2013; Abdul Aziz, Rokiah, & Ahmad Azrin, 2012; Chowdhury & Shaker, 2015; Zakir Hossain, 2009; Arif et al., 2012). Cambridge Dictionary (2019)defines Interest as money that is charged when you borrow money from a bank or money that is paid to you for the use of your money(Hossain, 2009).

Stance of bankers on interest prohibition has evolved substantially. Moreover, Shariah scholar's views about permissibility of banking have also changed. Use of profit and loss sharing mechanisms by managers at banks for earning profit is one evidence of the evolution. Moreover, Shariah scholars used to avoid all sorts of interest based or related tools and transactions. But now, shariah scholars allow interest based LIBOR (London Interbank Offer Rate) as a benchmark for calculatingcost of funds(Meera, Omar, & Noor, 2010;Al-wesabi, 2012). Nouman and Ullah (2018) report that Shariah scholars don't consider Murabaha as ideal mode of financing and have allowed its use only as a necessity.It seems to be happening as both perspectives are acknowledging each other value propositions and seems to be the result of sufficient knowledge diffusion.

Shariah compliance in banking cannot be claimed without avoiding interest based dealings. Therefore, SSB's established within IB's, ensure avoidance of interest in all sorts of dealings. The depositors are given permissible profit while interest free financing is providing to the businesses(Chong and Liu, 2009).

### 3.4.2.Uncertainty (Gharrar) prohibition

Gharrar has an Arabic root meaning 'absence' or 'insufficient information', which causes uncertainty(Kabir & Lewis, 2007). This may exist in a contract, price, subject or transaction outcome. There are degrees of Gharrar, and excessive Gharrar is prohibited in Shariah(Naughton, 2000). This is because excessive Gharrar leads to undue loss for one party and unjustified gain for the other. Gharrar may also be defined as any element of uncertainty, in any business orcontract, about the subject of the contract or its price which is forbidden in Islam( Obaidullah, 2005; Kabir & Lewis, 2007;Chowdhury & Shaker, 2015). This causes speculative risk, which leads to undue loss for one party and unjustified enrichment for the other (Ahmed, 2014). Islam declares all gains prohibited if they are earned on chances. Normally, permissible businesses also face uncertain situations but prohibitions come if uncertainty is excessive in nature. In a contract, both parties should exchange all relevant information they have and avoid deceiving each other. Islam gives the option of cancelling the contract to the weaker party which is not informed sufficiently by the other party and results in a loss. Gharrar also enters in a deal where two interdependent contracts like two sales in one agreement are made. Islam prohibits such agreements(Obaidullah, 2005).

In business dealings, Islam prohibits uncertainty. Any sale transaction with uncertainty is prohibited. For instance sale of a land is not allowed if the seller doesn't have a legal title and possession of it. Therefore, Islamic banks cannot deal in futures, swaps and forward contracts (Chong and Liu, 2009). The only two exceptions of future sales is Salam and Istisna which are allowed but have special conditions for it. (Usmani (2002) defines salam and Istisna as" Salam is a kind of sale in which payment is spot while the delivery of the good is deferred" and Istisna as "It refers to such sale in which commodity is transacted before it comes into existence. It is basically an order to manufacture"

Future sale is allowed only in salam and istisna and their use in islamic banking is increasing day by day. Earlier, shariah experts didn't allow any future sale but later through research and consensus, they allowed it. Similarly managers also welcomed these two exceptions and are using them in providing financing to the clients.

Diffusion seems to have happened as both managers and shariah experts are understanding and acknowledging each other value propositions and are devising ways where shariah compliance and bank's profitability are not compromised.

### 3.4.3. Profit and loss sharing

Islamic banks cannot deal in interest, therefore, for earning, they do business using models which give interest free income to the bank as well as depositors(Uppal & Mangla, 2014; Chowdhury & Shaker, 2015; Farag et al., 2018).Customer's deposits are taken on profit and loss basis which are onward invested in different permissible businesses. Return on those businesses is shared with the depositors(Naughton, 2000).Islamic banking products can be divided into types. Liability side or depository products and asset side products providing financing. One of the basic goals in developing both side products is to earn permissible income free of interest and speculations etc. SSBs ensure halal income through avoiding prohibited elements and doing business with firms which don't engage in alcohol, tobacco, pornography, movie making etc.(Kabir & Lewis, 2007; Chong & Liu, 2009;Ullah, 2012).

Conventional banking earns from the spread which is the difference of interest charged to the borrowers and the one given to the lenders (depositors). Since Islamic banks cannot earn through interest therefore, shariah scholars introduced the concept of profit and loss sharing which was initially considered impractical by the managers. Later on, islamic banking practices showed that without interest banking was possible following profit and loss sharing modes. Initially, products made by shariah scholars were not lacking market viability as they were only focusing shariah compliance but later on after acquiring banking knowledge, they managed to develop such products with the managers which were shariah compliant as well as marketable.

### 3.4.4. Liability side products

The products on liability side of lslamic banks are deposits on Modarbah basis, waadia basis and Interest free loans or qard e hasna(Obaidullah, 2005). Modarbah is a form of a partnership in which one party provides capitaland other party provides investment expertise and divides profit on predetermined ratio. So, modarbah based accounts are on profit and loss sharing while Waadia basis and Qarz E Hasna are non - profit and loss accounts. Some of them are not linked with a specific investment project and are called unrestricted modarbah account while some of them are linked with a specific investment called restricted modarbah account (Marwan *et al.*, 2016).

### 3.4.5. Asset side products

On assets side, the products are debt and equity based which are used for providing financing to the customers. Debt based products are sale based like Morabaha, Mosawamah, Salam and Istisna while rental based are Ijarah based products(Obaidullah, 2005). Murabaha is a type of sale in which seller gives the break of price into cost and profit to the buyer. Banks normally use deferred Murabaha which is credit sale and price is paid in installments. Sale in future date is prohibited in Islam but exception is provided in Salam and Istisna in which commodity is provided in future dates and price in full is paid at the time of sale(Usmani, 2002). The bank takes full risk of loss if the item destroys before being sold out to the customer. In Ijarah based financing, physical assets are leased out and rentals are earned on it. Since the assets are owned by the bank, therefore, bank bears the loss if any loss occurs to the asset(Marwan *et al.*, 2016).

Equity based financing products are musharika in which IB's take a share in equity of a business and share profit and loss with the partners(Marwan *et al.*, 2016). Profit earned from these activities can be higher and lower depending on performance of the businesses. Profit earned from these activities is shared with depositors on pre-determined ratio. Hence the rate of profit is not guaranteed and fixed and may fluctuate based on the actual performance of the business. IB's can only provide their services to businesses having their nature in compliance to shariah. IB's cannot cater the needs of businesses of Alcohol, Tobacco, speculative and pornography(Chong and Liu, 2009).

### 3.5. Banking Principles

A bank is a commercial entity that establishes with the aim of earning profit for its shareholders. A bank earns the profit while working as a financial intermediary(Obaidullah, 2005;Saunders, 2008) that borrows excess money of people and institutions in the form of deposits and in return, offers interest(Gouge, 1979; Llewellyn, 1999; Alexiou & Sofoklis, 2009). Technically, deposit is a loan which a bank borrows from the customer and is bound to return it along with accumulated interest. The bank further lends it to the people and institutions that have a shortage of capital on a higher interest rate(Marshall, 2013). The difference among the rate of interest borrowed and lended capital is called the banking spread. This spread is the primary source of income for banks (Idiab & Ahmad, 2011).

Banking is one of the important components of today's global financial system. It exists in almost every part of the world. It has a long history through which it evolved into existing form and will still be evolving further. Its history can be traced in the era of Babylonians, ancient China and Egypt. Banking evolved along with evolution of money(Sangale, 2013). Pharos in Egypt, Kublai Khan in China, Kings of France and England etc. developed their own currencies in the form of coins and paper money and compelled their people using them as a medium of exchange. Rulers used to earn from making coins as there always remained a difference in the value of coin and cost of production. This difference is called seigniorage. This practice still goes on and US government earns around 400 billion dollars every year as seigniorage (Wilk and Weatherford, 2006).

Today's banking industry is highly regulated. It is allowed to take deposits from the public and provide advances to other people. Regulators have developed the system of fractional reserve system for banks for leaving a fraction of deposit with the bank while rest of deposits is used for lending. Minimum capital requirements of the banks are guided by the Basel accords(Sangale et al., 2013).

While going through the literature about banking and its history, it was found that it operates around the following principles which inform its practices:

### 3.5.1. Interest based dealings

Banks are commercial entities and exist for a profit. They are integral part of global business. Banks are financial intermediaries, take deposit from customers and offers fixed return on it. The deposits are then provided to customers having need of capital on fixed return higher than what they offer to depositors. The basic contract of the bank with its customers is borrowing and lending(Heffernan, 2005). Bank earns through mobilization of deposits. Initially, deposits were used to be made for safe keeping with the banks. Later on, banks started lending those deposits against a fixed rate of interest. For encouraging deposits, they also started offering interests to the depositors. Banks lends through issuance of letter of credit, commercial papers, notes, bonds, short and long term financing depending on capital requirements (Jacobs, 2011). Beside borrowing and lending, banks also purchase bills before their due date on amount less than its face value(Sangale et al., 2013).

Deposits are the liabilities of the bank which it has to return while the credit provided to its customers earns interest and are called assets of the bank. Banks develop liability and asset side products based on fixed return or interest (Ball, 2012).

### 3.5.1.1 Liability Side Products

Liability side products are mainly divided into two types. Demand and time deposits. Demand deposits are paid in full or partial whenever customer demands for it. Demand deposits are again of two types. Current accounts or deposits which don't offer interest but rather offer free services like money transfer, free cheque books etc. Second type of demand deposits is saving accounts which offer interest but free services are normally absent.

Time deposits can only be withdrawn before maturity when customer bears a penalty on it, otherwise, such deposits are paid by the bank when the contractual time matures. Rate of interest on time deposits is considerably high (Obaidullah, 2005). Interest rate is higher if time duration of deposit is higher(Sangale et al., 2013).

### 3.5.1.2. Asset side products

People in need of capital approach banks for financing. Banks develop different products catering varying needs of customers for instance financing agriculture, working capital, machinery, building construction. Asset side products are also based on interest based lending. Beside loans, banks also provides lines of credit to customers which are also called running finance in which a customer is given a ceiling up to which it can withdraw any amount as a loan from the bank (Obaidullah, 2005).

Rate of interest charged on loan depends on the credibility of the borrower, duration of the loan, collateral, repayment mode and the type of business. Normally, banks provide loans of shorter duration, however, long term loans are also provided depending on favorable nature of above mentioned conditions (Sangale et al., 2013).

### 3.5.2. Profitability

Bank's major responsibilities are of two types. Firstly, they have to pay the deposit whenever they are demanded or on maturity and secondly, generating sufficient profit for their shareholders. Both ends can meet well if adequate deposits are mobilized and safe lending is made. Bank has to pay interest on deposits. It has to incur administrative costs of maintaining offices, paying taxes, salaries, equipment etc. So, the bank has to consider all such expenditures while deciding the interest to be charged to its borrowers. Otherwise, if proper costing is not carried out, despite recovery of loans along with interest on time, the bank may be in loss. Moreover, the bank has to pay a minimum interest rate to its borrowers while the lending rates ate also effected with the inter - bank rates, market forces so the bank has to keep the administrative cost within limits in order to generate profit for the shareholders (Sangale et al., 2013).

While lending, the bank must be very careful, otherwise, they may end up into non-performing loans which have to be written off. Thus, reducing the profit for the shareholders and increasing difficulty in paying back the depositor's money (Siddiqi, 1997). For lending, safety of the loan is very important. Bank must give loan to a customer who has the willingness and capacity of returning the loan along with interest. The bank must ensure that borrower uses lended money on thriving business generating sufficient profit enabling him to return the liabilities (Sangale et al., 2013).

For increasing the profit, a bank strives for giving minimum rate of interest to its depositors while charging higher rate on the credit given to the borrowers(Alexiou and Sofoklis, 2009). Banks take lending decisions with great care and always look for recovery. If they don't recover the lended money, the profit they give to their depositors and shareholders will fall. Normally, several banks exist and market forces decide the rates on which a bank borrows or lend money. Therefore, the decision of charging these rates is a very delicate one and the bank decides very cautiously (Bourke, 1989; Saunders &Cornett, 2008).

Islamic banks are also commercial entities and exist for a profit. If we check financial statements of islamic banks, they are also focusing on profit but yes in a shariah compliant way. Islam encourages circulation of money and earning profit but that must be through the ways commanded by God. Same is reflected in the practices of members of SSB and the policies they make for running a bank in a shariah compliant way.

### 3.5.3. Solvency& Liquidity

A bank lends the money which it borrows from customers while lend it to those who need it. If it fails to recover the lended money, then it will be capital deficient resulting in default at part of the bank. Therefore, it remains one of the core objectives of a bank to remain solvent. Global financial institutions like international monetary fund (IMF) helps the banks in checking and maintaining adequate solvencies (Marcia & Saunders, 2008; Alexiou & Sofoklis, 2009; Ball, 2012).

Business Dictionary (2015)defines liquidity as a measure of the extent to which a person or organization has cash to meet immediate and short term obligations or assets that can be converted to do this. A loan if repaid is considered to be a safe one but if it the recovery is late it may result into difficulties for the bank in paying its deposits. So, for remaining sufficiently liquid, recovery of loan is not enough but it will be well in time(Sangale et al., 2013). For keeping an effective check on the bank lending, central bank decides a cash margin which prevents the banks from lending all the money they have as deposits. Although, a bank has to pay interest on all the deposits while can take interest on it only from its borrowers. It also leaves some cash with the bank for those depositors that make withdrawals more frequently, while a bank cannot ask their borrowers to repay loans before the agreed time. This makes the liquidity management, one of the fundamental area, a bank has to observe for its survival (Aspachs, Nier, & Tiesset, 2005; Heffernan, 2005; Marcia & Saunders, 2008; Ball, 2012; Sangale et al., 2013).

Banking practices and rules and regulations revolve around these principles. Products are developed keeping in view the interest rate to be offered to the depositors and rates to be charged to lenders. Credit appraisal systems are developed for ensuring recovery of loan, thus improving solvency and liquidity of the bank. Optimizing all these factors results into profits for shareholders.

Like conventional banks, Islamic banks also strive to remaining liquid and solvent. But Shariah doesn't allow islamic banks the use of interest based instruments. Therefore, sukuks (islamic bonds) were developed for managing the liquidity requirements of islamic banks (Majid, 2003). AAOIFI has also issued liquidity standards for islamic banks and islamic banks in 45 countries are following various regulations for being profitable and shariah compliant (Thompson Reuters, 2018). These regulations and standards are developed by people having understanding of shariah and banking which may be the outcome of banking – shariah knowledge sufficiently diffused. These regulations are implemented by the BOD and SSB and there better understanding helps in making banking policies aiming at remaining solvent and liquid.

## 4. Theoretical Framework

A framework shows the important constructs and concepts of the study and that how a researcher visualizes the concepts and their interaction. It can also be termed as a researcher map of the phenomenon being investigated. These concepts may come from theory, personal experiences and objectives of the study (Miles & Huberman, 1994). Frameworks may be focusing micro or macro level phenomenon (Neuman, 2007). Based on this set of literature, below mentioned pre – empirical theoretical framework is developed:

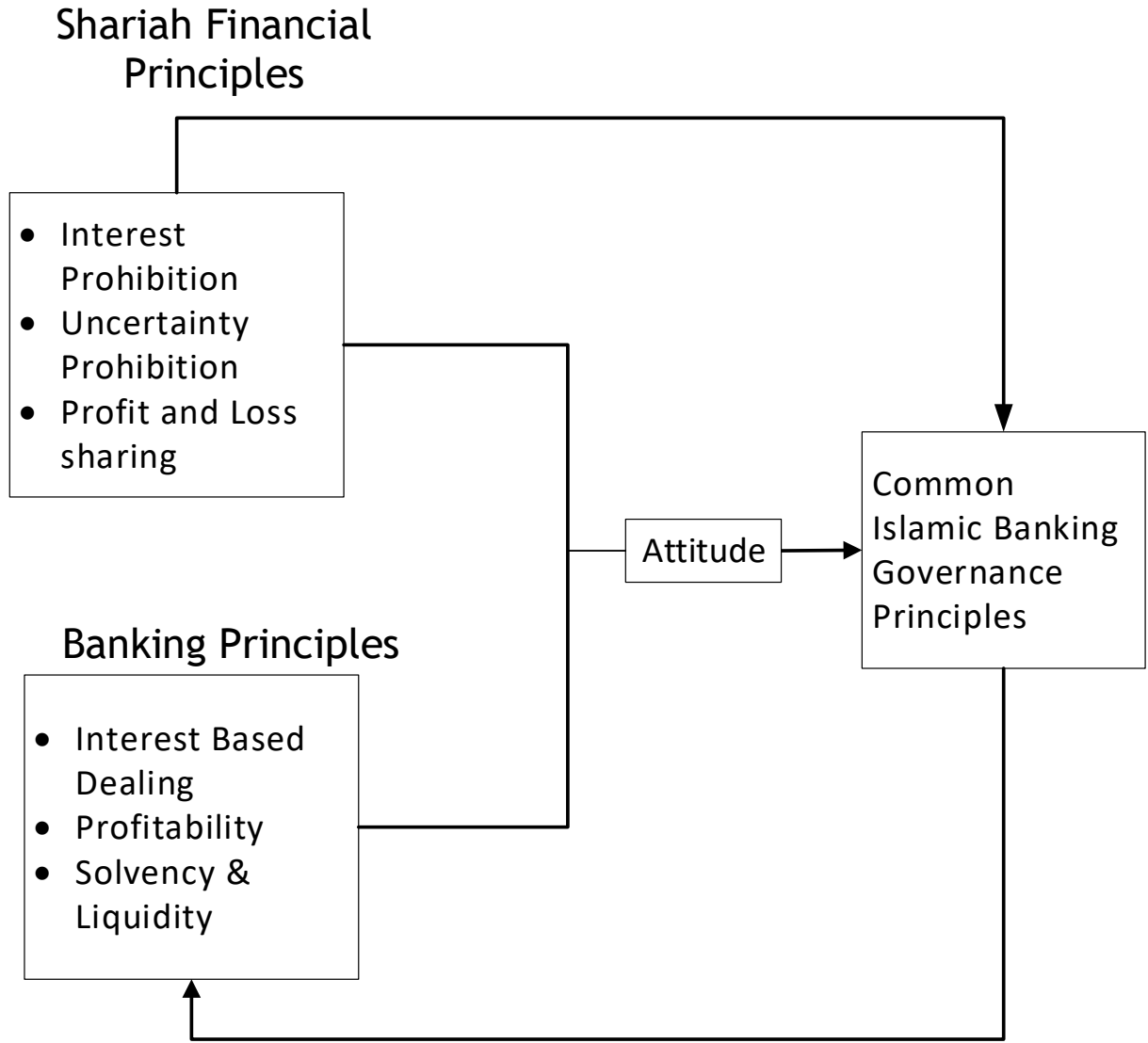

**Figure 1.1 Theoretical Frame work**

The above framework tries to say that shariah financial principles can be reduced to three principles informing the shariah governance practices in islamic banks namely; Interest Prohibition, Uncertainty Prohibition and Profit and loss sharing and banking principles informing governance practices can be reduced to Interest based dealing, Profitability, solvency and liquidity.

In the framework, the arrow coming from shariah financial principles towards common Islamic banking governance principles show that shariah informs the common Islamic banking governance principles, if any such principles exist. The arrow coming from common Islamic governance principles towards banking principles shows the common Islamic banking principles informing the banking principles.

Since, this study is more qualitative, therefore, with the data collection, if more principles emerge or if any principle is confirmed to be not relevant then the principles may be reduced or increased if they are identified by the participant shariah experts in the discipline. Presently, literature is guiding towards these principles.

Knowledge, attitude and Practice theory says that the practices of a person depends on the attitude developed through the knowledge, one receives. Banking education aims on generating profits for shareholders while optimizing the liquidity and remaining solvent based on interest based dealings. One internalizes the education through thinking and rationalizing. It starts impacting the way one thinks about generating profits, remaining liquid and solvent. After getting convinced with it, a person reflects on acquired knowledge and starts practicing it and develops a skill of implementing it. The practice of conventional banking focuses on interest based dealings generating profit while remaining sufficiently solvent and liquid that guide the banking practices and knowledge.

Similarly, shariah practices in Islamic banking governance may be informed by the principles of Interest Prohibition, Uncertainty Prohibition and Profit and loss sharing. Islamic banks also focuses on profit while remaining solvent and liquid but their dealings are free from interest and uncertainty and mainly rely on profit and loss sharing.

**5.     Conclusion**

This study has explored the literature on factors leading to governance of Islamic banks and has set the stage for further investigations that how cross diffusion of shariah and banking knowledge occurs among shariah scholars and managers working together which may be leading to emergence of new set of islamic banking governance principles. Moreover, this study has tried to take discussions towards the knowledge base of the hybrid people having understanding of shariah as well as banking working at governance level of islamic banks. Their knowledge and practices may be contributing into the emergence of common islamic banking governance principles. Data collected from such hybrid people may be adding new constructs to the developed literature based pre empirical framework and synthesis of post empirical framework describing and theorizing the knowledge diffusion process among managers and shariah scholars and exploring the governance principles of Islamic banks.

**References:**

AAOIFI (2018) *Accounting and Auditing Organization for Islamic Financial Institutions*. Available at: http://aaoifi.com/our-vision/?lang=en (Accessed: 16 August 2019).

Abdul Aziz, A., Rokiah, S. and Ahmad Azrin, A. (2012) 'Perception of Non-Muslims Customers towards Islamic Banks in Malaysia', *International Journal of Business and Social Science*, 3(11), pp. 151–163.

Ahmed, H. (2014) 'Islamic Banking and Shari ' ah Compliance : A Product Development Perspective', *Journal of Islamic Finance*, 3(2), pp. 15–29.

Al-wesabi, H. A. H. (2012) *Credit risk in islamic banks of GCC countries (Master dissertation)*. Universiti Utara Malaysia.

Alexiou, C. and Sofoklis, V. (2009) 'Determinants of bank profitability: evidence from the Greek banking sector', LIV(182), pp. 93–118. doi: 10.2298/EKA0982093A.

Arif, Muhammad; hussain, ashiq; azeem, M. (2012) 'University of Education', *International Journal of Humanities and Social Science*, 2(6), pp. 141–159.

Aspachs, O., Nier, E. and Tiesset, M. (2005) 'Liquidity , banking regulation and the macroeconomy Evidence on bank liquidity holdings from a panel of UK-resident banks', *Eelectronic copy downloaded from SSRN*, pp. 1–26.

Bashir, A.-H. M. (1999) 'Risk and Profitability Measures in Islamic Banks: The Case of Two Sudanese Banks', *Islamic Economic Studies*, 6(2), pp. 1–24.

Bhala, R. (2013) 'Overview of Islamic Finance', *Handbook of Key Global Financial Markets, Institutions and Infrastructure*, pp. 459–473. doi: 10.1016/B978-0-12-397873-8.00039-6.

Bourke, P. (1989) 'Concentration and other determinants of bank', *Journal of Banking and Finance*, 13, pp. 65–79.

Butt, S. A. (2010) *Corporate governance an introductory text for Pakistan*.

Chong, B. S. and Liu, M. H. (2009) 'Islamic banking: Interest-free or interest-based?', *Pacific Basin Finance Journal*. Elsevier B.V., 17(1), pp. 125–144. doi: 10.1016/j.pacfin.2007.12.003.

Chowdhury, N. T. and Shaker, & F. (2015) 'Shariah Governance Framework of the Islamic Banks in Malaysia', *International Journal of Management Sciences and Business Research.*, 4(10), pp. 115–124.

Dictionary, B. (2015) *Business Dictionary Liquidity*, *2015*. Available at:

http://www.businessdictionary.com/definition/liquidity.html (Accessed: 17 August 2019).

Dictionary, C. (2019) *Cambridge dictionary*. Available at: https://dictionary.cambridge.org/dictionary/english/interest (Accessed: 16 August 2019).

Elasrag, H. (2014) 'Corporate Governance in Islamic Finance Basic Cconcepts and Issues', *SSRN*, 244.

Farag, H., Mallin, C. and Ow-Yong, K. (2018) 'Corporate governance in Islamic banks: New insights for dual board structure and agency relationships', *Journal of International Financial Markets, Institutions and Money*, 54, pp. 59–77. doi: 10.1016/j.intfin.2017.08.002.

Farook, S. Z., & Farooq, M. O. (2011) 'Sharian Governance for Islamic Finance: Challenges and Pragematic Solutions', *Http://Ssrn. Com/Abstract= 1813483*.

Gouge, W. M. (1979) *A Short History of money & banking*. Augustus M. Kelley· Publishers New York New York. doi: 10.2307/4090985.

Grais, W. and Pellegrini, M. (2006) *Corporate Governance and Shariah Compliance in Institutions Offering Islamic Financial Services*. doi: http://dx.doi.org/10.1596/1813-9450-4052.

Grassa, R. (2013) 'Shariah supervisory system in Islamic financial institutions ', *Humanomics*, 29(4), pp. 333–348. doi: 10.1108/h-01-2013-0001.

Grassa, R. (2018) 'Shariah supervisory systems in Islamic finance institutions across the OIC member countries An investigation of regulatory frameworks', *Journal of Financial Regulation and Compliance*, (May). doi: 10.1108/JFRC-02-2014-0011.

Hamza, H. (2013) 'Sharia governance in Islamic banks : effectiveness and supervision model',

*International Journal of Islamic and Middle Eastern Finance and Management*, 6(3), pp. 226–237. doi: 10.1108/IMEFM-02-2013-0021.

Han, Q. (2010) *Practices and Principles in Service Design_ Stakeholder, Knowledge and and community of service*.

Haridan, N. M., Hassan, A. F. S. and Karbhari, Y. (2018) 'Governance , religious assurance and Islamic banks : Do Shariah boards effectively serve ?', *Journal of Management and Governance*. Springer US, 22(4), pp. 1015–1043. doi: 10.1007/s10997-018-9418-8.

Hassan, A. F. S. *et al.* (2017) 'Reporting assurance for religious compliance in islamic banks : Are we there yet ? Reporting assurance for religious compliance in islamic banks : are we there yet? Recent market surveys have indicated that Islamic banking assets account for', *International Journal of Applied Business and Economic Research*, 14(November), pp. 1465–1479.

Heffernan, S. (2005) *Money banking and Financial markets*. John Wiley & Sons.

Idiab & Ahmad (2011) 'Importance of commercial banks', *Journal of Applied Sciences Research*, 7(7), pp. 1127–1132.

IFSB (2009) *Guiding Principles on Shariah Governance systems for IFIs*. Available at: www.ifsb.org.

Ilieva, J., Ristovska, N. and Kozuharov, S. (2017) 'Banking Without Interest', *UTMS Journal of Economics*, 8(2), pp. 131–139.

Iqbal, Z. (1997) 'Islamic Financial Systems', *Finance & Development*, 34(June), pp. 42–45. Available at: https://www.imf.org/external/pubs/ft/fandd/1997/06/pdf/iqbal.pdf.

Jacobs, J. S. S. and H. (2011) *Handbook of Corporate Lending A Guide for Bankers and Financial Managers Table of Contents*.

Kabir & Lewis (2007) *Handbook of Islamic Banking*, *Handbook of Islamic Banking*. Available at: https://books.google.co.id/books?id=jvTtDzD5uFQC.

Kahf, M. (2006) 'Maqasid al Shari ' ah in the Prohibition of Riba and their Implications for Modern Islamic Finance', in *Presented Paper at IIUM International Conference on Maqasid Al-Shari'ah*. IIUM International Conference on Maqasid al Shari'ah.

Kasim, NuHtay, A. (2013) 'Comparative Analysis on AAOIFI , IFSB and BNM Shari ' ah Governance Faculty of Accountancy', *International Journal of Business and Social Science*, 4(15), pp. 220–227.

Kuran, T. (2004) *Islam and Mammon_ The Economic Predicaments of Islamism - Timur Kuran - Google Books*. Princeton University Press.

Llewellyn, D. T. (1999) *The New Economics of Banking*.

Majdoub, J. (2015) 'How ethical is islamic banking in the light of the objectives of islamic law?', *Journal of Religious Ethics*, pp. 51–77.

Majid, A. R. A. (2003) 'Development of Liquidity Management Instruments: Challenges and Opportunities', in *International Conference on Islamic Banking: Risk Management, Regulation and Supervision - 2003*, pp. 1–24. Available at: http://www.sbp.org.pk/departments/ibd/Lecture_6_LIQUIDITY_MANAGEMENT.pdf.

Marshall, A. (2013) *Principles of Economics*. 8th Editio. Palgrave Macmillan. doi: 10.1007/978-

1-137-37526-1.

Marwan, M. *et al.* (2016) 'The importance of the shari ' ah supervisory boards ( ssbs ) in the islamic banks', *South East Asia Journal of Contemporary Business, Economics and Law*, 9(2), pp. 25–31.

Matthew B. Miles, A. M. H. (1994) *Qualitative Data Analysis*. SAGE Publications, Inc.

Meera, A. K. M., Omar, M. A. and Noor, A. M. (2010) 'An Islamic pricing benchmark', *ISRA Research Paper*, (17), pp. 1–78. Available at: http://irep.iium.edu.my/16770/.

Mehtab, H., Zaheer, Z. and Ali, S. (2015) 'Knowledge, Attitudes and Practices (KAP) Survey: A Case Study on Islamic Banking at Peshawar, Pakistan', *FWU Journal of Social Sciences*, 9(2), pp. 1–13. Availableat: https://search.proquest.com/docview/1761253560?accountid=14511%0Ahttp://sfx.ucl.ac.uk/sfx_local?url_ver=Z39.88-2004&rft_val_fmt=info:ofi/fmt:kev:mtx:journal&genre=article&sid=ProQ:ProQ%253Acentral&atitle=Knowledge%252C+Attitudes+and+Practices+%2528KAP%252.

Naughton, S. N. and T. (2000) 'Religion, ethics and stock trading: The case of an Islamic equities market', *Journal of Business Ethics*, 23(2), p. 145. Available at: http://proquest.umi.com/pqdlink?did=49312017&Fmt=7&clientId=13939&RQT=309&VName=PQD.

Neuman, W. L. (2007) *Basics of Social Research: Qualitative and Quantitative Approaches*. SAGE Publications, Inc.

Nouman, M. and Ullah, K. (2018) 'Why Islamic Banks Tend to Avoid Participatory Financing ? A Demand , Regulation , and Uncertainty Framework', *Business & Economic Review*, Vol. 10(June), pp. 1–32. doi: 10.22547/BER/10.1.1.

Obaidullah, M. (2005) 'Islamic Financial Services'. King Abdul Azziz University, Saudi Arabia, p. 282.

Othman, A. A. *et al.* (2013) 'Shariah governance for Islamic financial institutions in Malaysia on the independency of Shariah committee and efficiency on its Shariah decisions', *7th WSEAS International Conference on Management, Marketing and Finance (MMF '13)*, pp. 93–100. Available at: http://irep.iium.edu.my/38380/.

Othman, R. and Mara, U. T. (2013) 'Islamic Banking Products : Regulations , Issues and Challenges', *The Journal of Applied Business Research*, 29(4), pp. 1145–1156. doi: 10.19030/jabr.v29i4.7922.

Qorchi, M. El (2005) 'Islamic Finance Gears Up *', *Finance and Development, , 46.*, p. 46.

Rammal, H. G. (2015) 'Islamic finance: Challenges and opportunities', *Journal of Financial Services Marketing*, 15(2010), pp. 189–190. doi: 10.1057/fsm.2010.15.

Report, G. I. F. (2019) *Global Islamic Finance*.

Reuters, T. (2018) *Islamic Finance Development Report 2018 Building Momentum*.

Rogers, E. M. (2010) 'Diffusion of Innovations', in *Diffusion of Innovations*. Simon and Schuster Inc, p. 35.

S&P (2018) *Islamic Finance Outlook*.

Sangale et al. (2013) *Fundamentals of banking*. University of Pune, India.

Saunders, M. & (2008) *Financial Institutions Management*.

Siddiqi, H. A. (1997) *Practice and Law of Banking in Pakistan*. Royal Book Co: Karachi.

Siddiqi, M. N. (2006) 'Shariah , Economics and the Progress of Islamic Finance : The Role of Shariah Experts', in *Seventh Harvard forum on Islamic finance Cambridge, Massachusetts, USA*, pp. 1–7.

Siddique, Z. (2015) 'Islamic Economics: A Plea for Islamic Capitalism', *SSRN Electronic Journal*, Vol. 13, N(02). doi: 10.2139/ssrn.2710151.

Ullah, Harwood and Jamali (2016) '" Fatwa Repositioning ": The Hidden Struggle for Shari ' a Compliance Within Islamic Financial Institutions', *Journal of Business Ethics*. Springer Netherlands, 149(4), pp. 895–917. doi: 10.1007/s10551-016-3090-1.

Ullah, K. and Patel, N. V (2011) 'Addressing Emergent Context of Shariah Compliant Financial Services : A Service Designing Construct', *International Review of Business Research Papers*, 7(3), pp. 81–93.

Ullah, S. (2012) *Fatwa Repositioning: the hidden struggle for shariah compliance within Islamic Financial Institutions*, *PhD Thesis.* University of Southampton, UK.

Uppal, J. Y. and Mangla, I. U. (2014) 'Islamic Banking and Finance Revisited after Forty Years: Some Global Challenges', *Journal of Finance*, (January 2014).

Usmani, I. (2002) *Meezan Bank's Guide to Islamic Banking*. Darul - ishaat urdu bazar Karachi, Pakistan.

Wajdi Dusuki, A. (2008) 'Understanding the objectives of Islamic banking: a survey of stakeholders' perspectives', *International Journal of Islamic and Middle Eastern Finance and Management*, 1(2), pp. 132–148. doi: 10.1108/17538390810880982.

Webster, M. (2019) 'Definition of diffusion 1 ':Wilk, R. and Weatherford, J. (2006) *The History of Money: From Sandstone to Cyberspace.*, *The Journal of the Royal Anthropological Institute*. Crow n Publishers , Inc . • Ne w Yor k. doi: 10.2307/3034519.

Wilson (1999) 'Challenges and Opportunities for Islamic Banking and Finance in the West : The United Kingdom Experience', *Thunderbird International Business Review*, 41(October), pp. 421–444.

Wilson, R. (2000) 'Challenges and opportunities for islamic banking and finance in the west : the United Kingdom experience', 7.

Yu, F. (2015) 'Frank Yu and Xiaoyun Yu * This Version : June 2010 Journal of Financial and Quantitative Analysis forthcoming', *Journal of Financial and Quantitative Analysis*, (April).

Yusoff, M. B. (2013) 'Riba, profit rate, islamic rate and market equilibrium', *International Journal of Economics,Management and Accounting*, 1(1), pp. 33–63.

Zahid, S. N. and Khan, I. (2019) 'Islamic Corporate Governance: The Significance and Functioning of Shari'ah Supervisory Board in Islamic Banking', *Turkish Journal of Islamic Economics*, 6(1), pp. 87–108. doi: 10.26414/A048.

Zakir Hossain, M. (2009) 'Why is interest prohibited in Islam? A statistical justification', *Humanomics*, 25(4), pp. 241–253. doi: 10.1108/08288660910997610.